# Noise of a chargeless Fermi liquid

Cătălin Pașcu Moca,[1, 2] Christophe Mora,[3] Ireneusz Weymann,[4] and Gergely Zaránd[1]

[1]*BME-MTA Exotic Quantum Phase Group, Institute of Physics,
Budapest University of Technology and Economics, H-1521 Budapest, Hungary*
[2]*Department of Physics, University of Oradea, 410087, Oradea, Romania*
[3]*Laboratoire Pierre Aigrain, École Normale Supérieure,
Université Paris 7 Diderot, CNRS; 24 rue Lhomond, 75005 Paris, France*
[4]*Faculty of Physics, Adam Mickiewicz University, 61-614 Poznań, Poland*
(Dated: January 22, 2017)

We construct a Fermi liquid theory to describe transport in a superconductor-quantum dot-normal metal junction close to the singlet-doublet (parity changing) transition of the dot. Though quasiparticles do not have a definite charge in this chargeless Fermi liquid, in case of particle-hole symmetry, a mapping to the Anderson model unveils a hidden U(1) symmetry and a corresponding pseudo-charge. In contrast to other correlated Fermi-liquids, the back scattering noise reveals an effective charge equal to the charge of Cooper pairs, $e^* = 2e$. In addition, we find a strong suppression of noise when the linear conductance is unitary, even for its non-linear part.

PACS numbers: 72.10.-d, 73.21.La, 74.45.+c

*Introduction.–* Apart from disrupting our communication, the noise contains interesting and abundant information. Advances in experimental techniques gradually allowed us to enter the quantum regime, and access this information through noise measurements in various setups, ranging from nano-circuits [1–6] or quantum optics devices [7, 8] to bosonic and superfluid systems in cold atomic settings [9, 10]. Its structure reveals the quantum statistics and the charge of quasiparticles as well as the nature of their interactions. For example, bunching of photons in quantum optics reflects the bosonic nature of light [11], while the complete suppression of noise in case of a perfectly transmitting conductance channel in a nano-circuit [12] is a consequence of the electrons being fermions.

Low temperature noise has been used to extract the transmission amplitudes of a point contact [5, 13] but also to gain insight into the structure of quantum fluctuations [1, 14]. In such interacting nano-contacts, the shot noise carries informations on the structure of residual interactions and elementary excitations. For example, the noise of back-scattered current in quantum-Hall devices has been used, e.g., to extract the fractional charge $e^*$ of excitations at fillings $\nu = 1/3$ [15, 16] or $\nu = 2/3$ [17]. Furthermore, in a strongly interacting local Fermi liquid, realized, e.g., in a quantum dot (QD) attached to normal electrodes at very low temperatures, the noise of the back-scattered current is induced by interactions, and the corresponding effective charge, $e^* = 5e/3$ turns out to reflect the structure of local interactions rather than that of elementary excitations [4, 18, 19].

Superconductors, from this perspective, are of particular interest; while they obviously carry current, elementary excitations in a superconductor do not have a definite charge, and possess only spin. In particular, attaching a superconductor to normal electrodes destroys the charge of electrons in its neighborhood by proximity effect [20], and makes charge ill-defined, there too. Here we wish to understand the structure of low temperature electric noise of these *chargeless* excitations in the presence of strong interactions. For this purpose, we propose to study the superconductor-quantum dot-normal metal (S-QD-N) system depicted in Fig. 1, and investigated experimentally by several groups [21–25]. The set-up consists of an artificial atom attached to a superconductor, and probed by a weakly coupled normal electrode or a scanning tunneling microscope (STM) tip.

As we discuss below, at very low temperatures, this system becomes a Fermi liquid of chargeless quasiparticles. Nevertheless, we can still describe it in terms of Noziéres' Fermi liquid theory [26], generalized for the Anderson model [27]. In the limit of maximal conductance, the current through this device turns out to be almost 'noiseless', surprisingly up to (and including) $\mathcal{O}(V^3)$ order. Moving slightly away from this sweet point, the shot noise $S$ at zero temperature is found to be generated by the back-scattering current $\delta I$ [18, 28, 29], yielding an effective charge,

$$e^* = \frac{S}{\delta I} = 2e + \mathcal{O}(eV)^3 \,. \qquad (1)$$

In stark contrast to regular Fermi liquids [18, 30], this charge appears to be related only to the fact that electrons enter the superconductor as Cooper pairs and, surprisingly, is not influenced by the otherwise strong interactions, a prediction that could easily be verified experimentally.

*Effective field theory.–* Throughout this work, we shall focus on the local moment regime, where there is just a single electron on the QD, which can be described as a single spin $S = 1/2$, coupled antiferromagnetically to the two electrodes by an exchange coupling (see Fig. 1). In the absence of the normal electrode, the spin on the dot binds a quasiparticle to itself antiferromagnetically, appearing as an excited singlet state $|0\rangle$ inside the gap [31] when the exchange coupling $\mathcal{J}$ is weak. With increasing $\mathcal{J}$, the energy of this so-called Shiba state drifts towards zero, until at a critical value $\mathcal{J} = \mathcal{J}_c$ it becomes smaller than that of the doubly degenerate spin states (dressed by the superconductor), $|\pm\rangle$, and a parity changing phase transition occurs [21, 24, 32–35].

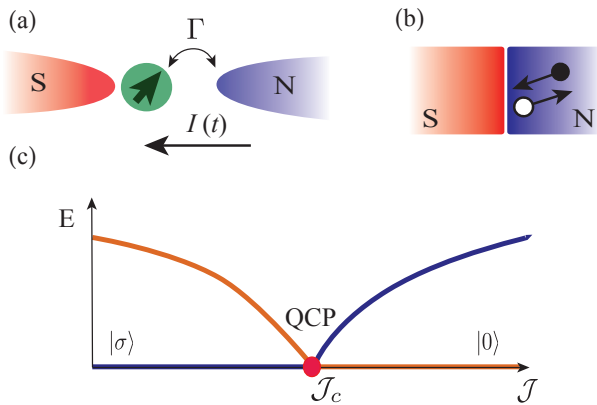

FIG. 1. (Color online) (a) A localized spin in a quantum dot is coupled to a superconductor (S) and weakly coupled to a normal (N) lead, too. (b) Typical Andreev scattering processes in the normal lead: One electron is reflected back as a hole, while a Cooper pair is transmitted to the superconductor-quantum dot device. (c) Sketch of the subgap states in the S-QD system, in the absence of the metallic lead. At $\mathcal{J} = \mathcal{J}_c$ a parity-changing transition occurs. For $\mathcal{J} > \mathcal{J}_c$ the ground state is a many-body singlet $|0\rangle$, while in the opposite limit it is a doublet $|\sigma\rangle$.

Coupling the QD weakly to a normal electrode at $\mathcal{J} \approx \mathcal{J}_c$ induces strong quantum fluctuations between the states $|\pm\rangle$ and $|0\rangle$, and leads to a strongly correlated local Fermi liquid state [26, 36, 37] with close to perfect transmission and a conductance [38] $G \approx 4e^2/h$. Though our conclusions will turn out to be quite general, throughout this letter, we focus our attention on the vicinity of this transition, where tunneling between the QD and the normal electrode can be described in terms of the three states, $|0\rangle$ and $|\sigma\rangle$, and the following simple Hamiltonian,

$$H_v = \sum_\sigma \left[ |\sigma\rangle\langle 0| \left(v_\sigma^+ \psi_\sigma - v_{\bar\sigma}^- \psi_{\bar\sigma}^\dagger\right) + h.c.\right]. \quad (2)$$

Here $v_\sigma^+ \propto \langle\sigma|d_\sigma^\dagger|0\rangle$ and $v_{\bar\sigma}^- \propto \langle\sigma|d_{\bar\sigma}|0\rangle$ [39], and the field $\psi_\sigma^\dagger$ denotes the creation operator of an electron in the normal electrode. Thus the total Hamiltonian reads

$$H = H_v + \Delta E \left( \sum_\sigma |\sigma\rangle\langle\sigma| - |0\rangle\langle 0| \right) + H_\psi \quad (3)$$

with the free electron Hamiltonian $H_\psi$ generating the dynamics of the field $\psi$ and the term $\sim \Delta E$ controlling the parity-changing transition.

Electron-hole symmetry and time reversal symmetry imply the relations $v_\uparrow^+ = v_\uparrow^- = -v_\downarrow^- = v$ for the tunneling matrix elements [40], and allow one to further simplify the problem by expressing the fields $\psi_\sigma$ in terms of new fields, $\Phi_\sigma$ through the expressions $\psi_{\uparrow/\downarrow} \equiv (\Phi_{\uparrow/\downarrow} \mp \Phi_{\downarrow/\uparrow}^\dagger)/\sqrt{2}$, and thereby mapping the problem to the mixed valence Anderson model [41]. Notice, that the new fields $\Phi_\sigma$ contain both electron and hole excitations, and have no charge. Nevertheless, in terms of these, the Hamiltonian (3) possesses a residual U(1) symmetry, generated by global gauge transformations, $\phi_\sigma \to e^{i\alpha}\phi_\sigma$, $|\pm\rangle \to e^{i\alpha}|\pm\rangle$. The corresponding 'pseudocharge' is a remainder of a hidden charge SU(2) symmetry, associated with electron-hole symmetry, broken down to U(1) by the presence of superconductivity [42].

*Fermi liquid theory.-* Following Landau and Noziéres, we now construct a Fermi liquid theory to study non-equilibrium transport in this system [26]. This is most easily achieved in terms of the field $\Phi_\sigma(x) = \int d\epsilon\, \varphi_{\epsilon\sigma}(x)\, b_{\epsilon\sigma}$, which we now expand in the basis of incoming chiral scattering states, $\varphi_{\epsilon\sigma}(x)$, and the corresponding annihilation operators, $b_{\epsilon\sigma}$. Notice that the $x < 0$ and $x > 0$ parts of these states correspond to incoming and reflected plane waves in the normal electrode, and behave as $\sim e^{ikx}$ and $\sim S_\sigma e^{ikx}$, respectively, with $k = \epsilon/v_F$ and $S_\sigma = e^{2i\delta_\sigma}$ denoting the scattering amplitude.

In terms of the field $\Phi_\sigma$, the effective Hamiltonian is just that of the Anderson model in the limit of infinite local interaction $U$, i.e., with forbidden double occupancy. Therefore the ground state is a Fermi liquid which, at low temperatures, can be described in terms of a local field theory, containing the field $\Phi_\sigma(x)$ and its derivatives at $x = 0$ [43, 44]. The structure of this Hamiltonian is simply dictated by symmetries: it must conserve spin and the hidden U(1) charge, however, unlike Noziéres' original theory [26], it is not 'electron-hole' symmetrical in terms of the hidden U(1) charge [43]. In terms of the quasiparticle operators $b_{\epsilon\sigma}$, this effective Hamiltonian thus reads

$$H_{\rm FL} = \sum_\sigma \int_\varepsilon \varepsilon\, b_{\varepsilon\sigma}^\dagger b_{\varepsilon\sigma} + H_\alpha + H_\phi + \ldots \quad (4)$$

$$H_\alpha = -\sum_\sigma \int_{\varepsilon_1,\varepsilon_2} \left[\frac{\alpha_1}{2\pi}(\varepsilon_1+\varepsilon_2) + \frac{\alpha_2}{4\pi}(\varepsilon_1+\varepsilon_2)^2\right] b_{\varepsilon_1\sigma}^\dagger b_{\varepsilon_2\sigma}$$

$$H_\phi = \int_{\varepsilon_1,\ldots,\varepsilon_4} \left[\frac{\phi_1}{\pi} + \frac{\phi_2}{4\pi}(\sum_{i=1}^4 \varepsilon_i)\right] : b_{\varepsilon_1\uparrow}^\dagger b_{\varepsilon_2\uparrow} b_{\varepsilon_3\downarrow}^\dagger b_{\varepsilon_4\downarrow} :,$$

where $:\cdots:$ denotes the normal ordering, and the coefficients can be expressed as universal functions of the ratio $\Delta E/\Gamma$ with the hybridization $\Gamma \sim v^2$ (see Supplemental Material [45]).

*Current and noise.–* Our main purpose is to determine the expectation value and the noise of the current $\hat{J}(r) = ev_F \sum_\sigma (\psi_\sigma^\dagger(-r)\psi_\sigma(-r) - \psi_\sigma^\dagger(r)\psi_\sigma(r))$ at position $r > 0$, in the limit of low temperatures and small bias voltages. Introducing the auxiliary quasiparticle fields $b_\sigma(x) \equiv \int d\epsilon\, e^{ikx}\, b_{\epsilon\sigma}$, and exploiting the asymptotic behavior of the scattering states for a position sufficiently far away from the scattering region, we find

$$\hat{I} = ev_F \left[b_\downarrow(-r)b_\uparrow(-r) - S_\downarrow b_\downarrow(r) S_\uparrow b_\uparrow(r) + {\rm h.c.}\right], \quad (5)$$

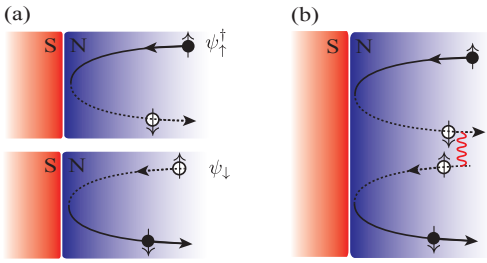

FIG. 2. (Color online) (a) The "charged" quasiparticles $a_{\epsilon\sigma}$ correspond to incoming electrons reflected as holes. Each unscattered quasiparticle therefore transfers a charge $2e$ to the superconducting side. (b) A potential scattering event $a^{\dagger}_{\varepsilon\uparrow}|\text{FS}\rangle \to a_{-\varepsilon\downarrow}|\text{FS}\rangle$ creates an outgoing electron state, i.e., a reflected charge, and reduces the current by $2e$.

where for the sake of generality we kept the spin dependence of the scattering matrix, relevant in case of external magnetic fields. Representing incoming scattering states, the expectation values of the products of the operators $b_{\varepsilon\sigma}$ are then determined by the normal electrodes (see Supplemental Information [45]), enabling us to compute physical quantities such as $\langle \hat{I} \rangle$ or its correlation functions perturbatively in the interaction terms of Eq. (4).

In the following, for the sake of simplicity, we shall focus on the non-equilibrium regime where the temperature is much smaller than the applied voltage bias, $eV \gg T \approx 0$. There the zero'th order expectation value of the current, e.g. simply yields $\langle \hat{I} \rangle = G_0 V + \ldots$ where the linear conductance recovers the known expression [12]

$$G_0 = \frac{2e^2}{h}\big(1 - \Re\mathfrak{e}(S_{\downarrow}S_{\uparrow})\big) = \frac{4e^2}{h}\sin^2(\delta_{\uparrow}+\delta_{\downarrow})\,, \quad (6)$$

with $\delta_\sigma$ the phase shifts of the (chargeless) quasiparticles at the Fermi energy, $S_\sigma = e^{2i\delta_\sigma}$. In the so-called unitary limit, $\delta_\sigma \to \delta_0 = \pi/4$ [37], corresponding to a precisely half-filled level, the conductance takes its maximal value, $G_0 = 4e^2/h$. Back to the original model, the corresponding ground state is an equal superposition of the singlet and doublet states. Here, every incoming electron in the normal electrode is perfectly reflected as a hole, and the current turns out to be noiseless (at least up to third order in the voltage $V$), similar to a perfectly transmitting point contact [46]. To observe the charge of the carriers, we therefore need to move slightly away from this unitary limit by setting $\delta = \pi/4 + \tilde{\delta}$ and, similar to Refs. [17, 18, 29] focus on *backscattered* current, defined as the correction with respect to the maximal current, $\hat{I} = I_u - \delta\hat{I}$.

At the level of the effective field theory, we can move away from the unitary limit by just adding a term $\widetilde{H}_\delta = -(\tilde{\delta}/\pi)\sum_\sigma \int_{\varepsilon,\varepsilon'} b^{\dagger}_{\varepsilon\sigma}b_{\varepsilon'\sigma}$ to the unitary Hamiltonian, and treating its effect perturbatively on the unitarily scattered quasiparticles, similarly to the other terms in Eq. (4). To see how these processes generate the back-scattered current, we rewrite $\widetilde{H}_\delta$ in terms of "charged" unitary quasiparticle operators, $a_{\epsilon\uparrow/\downarrow} = (b_{\epsilon\uparrow/\downarrow} \mp b^{\dagger}_{-\epsilon\downarrow/\uparrow})/\sqrt{2}$ as

$$\widetilde{H}_\delta = -(\tilde{\delta}/\pi)\sum_\sigma \int_{\varepsilon,\varepsilon'}(a^{\dagger}_{\varepsilon\uparrow}a^{\dagger}_{\varepsilon'\downarrow} + \text{h.c.}).$$

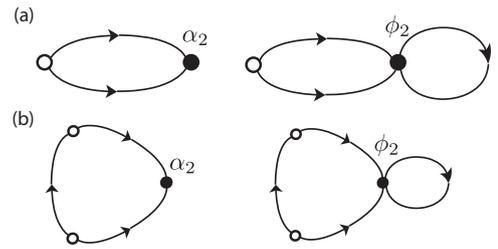

FIG. 3. First order diagrams describing the corrections to the current (a) and to the noise (b). The open circles represent the current vertices while the filled dots corresponds to the interaction vertices. Only the non-vanishing contributions are displayed.

Here the word "charged" must be used with caution: the operator $a^{\dagger}_{\varepsilon\uparrow}$ creates namely a scattering state, which at time $t \to -\infty$ represents an incoming electron of charge $e$ on the normal side, while for $t \to \infty$ it is a reflected hole of charge $-e$. Scattering events generated by the term $\widetilde{H}_\delta$, e.g., convert an incoming electron state $a^{\dagger}_{\varepsilon\uparrow}|\text{FS}\rangle$ into an incoming hole state $a_{-\varepsilon\downarrow}|\text{FS}\rangle$ which, for time $t \to \infty$ represents a reflected *electron*. These reflected electrons *reduce* the transmitted current, and contribute to the back scattered current, $\delta\hat{I}$, as we schematically represent in Fig. 2.

The rate of these back-scattering events can be simply computed by applying Fermi's golden rule as

$$\Gamma_\delta = \sum_\sigma \left(\frac{2\pi}{\hbar}\right)\int_{\varepsilon_1\varepsilon_2}|\langle a^{\dagger}_{\varepsilon_2,-\sigma}H_\delta a^{\dagger}_{\varepsilon_1\sigma}\rangle|^2\delta(\varepsilon_1+\varepsilon_2),\quad (7)$$

yielding $\Gamma_\delta = 8\tilde{\delta}^2\, eV/h$, and a backscattering current $\delta I_\delta = 2e\Gamma_\delta$. Computing the rates of all scattering processes perturbatively we arrive at

$$\langle \delta\hat{I}\rangle = \frac{8e^2}{h}\,V\left\{2\tilde{\delta}^2 + \big(\frac{4}{3}\alpha_2 - \phi_2\big)\tilde{\delta}\,(eV)^2 + \ldots\right\}. \quad (8)$$

All processes that give contribution to $\delta I$ to this order turn out to be such that an incoming electron does not transfer charge to the superconductor, but is reflected back as an electron or as an electron and an electron-hole pair. Under the assumption that these are all independent Poissonian processes that reduce the current by $2e$, we immediately arrive at the conclusion, that the shot noise is just $2e\,\delta I$, yielding the effective charge (1).

The previous result can be generalized for arbitrary values of $\delta$ by incorporating the phase shift $\delta$ in the definition of the quasiparticle operators $a_{\epsilon\sigma}$ and the current operator, and then using the Keldysh approach to perform perturbation theory in the Fermi liquid coefficients [45]. The leading corrections to the expectation value of the current are shown as an example in Fig. 3, and yield

$$\langle \hat{I} \rangle = \frac{2e^2}{h} V \left\{ 2\sin^2 2\delta - \frac{\pi \chi_c'}{3} \sin 4\delta (eV)^2 + \dots \right\}, \quad (9)$$

where we made use of a Fermi liquid relation relating the derivative of the charge susceptibility of the effective Anderson model [27] to the Fermi liquid coefficients as, $\frac{4}{3}\alpha_2 - \phi_2 = -\pi\chi_c'/3$. By expanding this formula around $\delta_0 = \pi/4$, we recover Eq. (8).

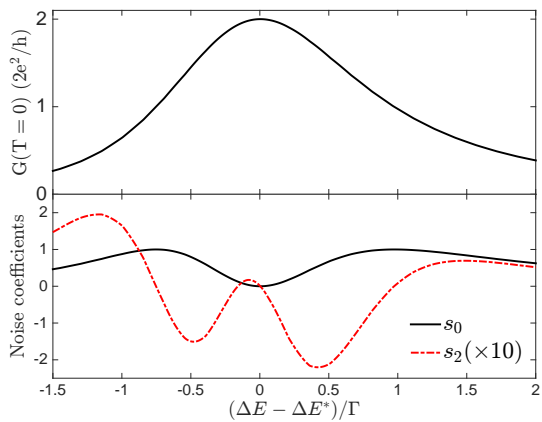

FIG. 4. (Color online) Dimensionless conductance (top panel) and the first non-vanishing dimensionless noise coefficients, $s_0$ and $s_2$ (bottom panel). The coefficient $s_2$ is magnified 10 times. The noise coefficients vanish at the point of maximal conductance, $\Delta E = \Delta E^*$, where $\delta = \delta_0 = \pi/4$.

The shot noise can be computed along similar lines [45]. It can be expressed as a power series in the voltage,

$$S(V, \Delta E) = e \frac{2e^2}{h} V \sum_{n=0}^{\infty} s_n(\Delta E)\, (eV/\Gamma)^n \quad (10)$$

with the first few dimensionless coefficients given by $s_0 = \sin^2 4\delta$, $s_1 = 0$, and $s_2 = -\Gamma^2 \frac{\pi \chi_c'}{3} \sin 8\delta$. Remarkably, at the point of maximal conductance, $\delta = \delta_0 = \pi/4$, both $s_0$ and $s_2$ vanish, and the current remains noiseless up to $\sim (eV/\Gamma)^4$ even in the presence of interactions (see Fig. 4). Similar to the dimensionless linear conductance, $g \equiv G/G_0 = \sin^2(2\delta)$, the dimensionless coefficients $s_n$ are universal functions of the ratio $\Delta E/\Gamma$. These functions can all be determined using Bethe Ansatz or numerical renormalization group calculations [27], and we have displayed them in Fig. 4.

We are now in the position of expressing the effective charge $e^*$ as a function of $\Delta E$ and $V$,

$$\frac{e^*}{e} \equiv \frac{S}{e\,\delta I} = \frac{\sin^2 4\delta - \frac{\pi\chi_c'}{3}\sin 8\delta (eV)^2 + \dots}{2\cos^2 2\delta + \frac{\pi\chi_c'}{3}\sin 4\delta (eV)^2 + \dots}. \quad (11)$$

For small voltages, this formula yields an effective charge $e^* = 2e(1-\delta g)$, while for larger voltages this value crosses over to $e^* = 2e(1-2\delta g)$, with $\delta g = 1-\sin^2(2\delta) = \cos^2(2\delta)$ representing the reduction of the $T=0$ temperature dimensionless linear conductance with respect to its maximal, unitary value. Unlike a usual, strongly interacting Fermi liquid, the effective charge remains close to $2e$ in the vicinity of the unitary scattering regime, even in the linear voltage regime. The effective charge of the chargeless Fermi liquid is thus that of Cooper pairs $e^* \approx 2e$, and it only slightly deviates from this value in spite of strong electron-electron interactions.

Our results are robust, and barely influenced by other external perturbations, too. An external magnetic field, e.g., splits up the phase shifts $\delta_{\uparrow,\downarrow}$. However, the expressions above depend only on the sum, $\delta_\uparrow + \delta_\downarrow$, and are therefore independent of the applied field to the order discussed here.

*Conclusions.–* In conclusion, we have determined the current noise of a S-QD-N device and have shown that, at low temperatures, transport can be described in terms of a chargeless Fermi liquid theory with a hidden pseudo-charge. At resonance, in spite of the presence of strong interactions and inelastic processes, the current remains noiseless up to $\mathcal{O}(V^3)$ order. Slightly off resonance, elastic and inelastic reflections of quasiparticles at the interface dominate the backscattered current and the associated current noise – the latter displaying universal behavior in the vicinity of the Shiba transition. Consequently, the effective charge remains constant $e^* \approx 2\,e$ close to resonant transmission, even in the non-linear noise regime.

*Acknowledgements.* We would like to thank Pascal Simon for important and fruitful discussions. This work has been supported by the Hungarian research grant OTKA K105149, UEFISCDI Romanian Grant No. PN-II-RU-TE-2014-4-0432 and the Polish National Science Centre Grant No. DEC-2013/10/E/ST3/00213.

# Supplementary material for the paper "Noise of a chargeless Fermi liquid"

Cătălin Pașcu Moca,[1,2] Christophe Mora,[3] Ireneusz Weymann,[4] and Gergely Zaránd[1]

[1]*BME-MTA Exotic Quantum Phase Group, Institute of Physics,*
*Budapest University of Technology and Economics, H-1521 Budapest, Hungary*
[2]*Department of Physics, University of Oradea, 410087, Oradea, Romania*
[3]*Laboratoire Pierre Aigrain, École Normale Supérieure,*
*Université Paris 7 Diderot, CNRS; 24 rue Lhomond, 75005 Paris, France*
[4]*Faculty of Physics, Adam Mickiewicz University, 61-614 Poznań, Poland*
(Dated: January 22, 2017)

## S-I. PERTURBATION THEORY ON THE KELDYSH CONTOUR

In this section we outline the calculation of the current using the perturbative approach on the Keldysh contour [1]. The expression for the current operator at a distance $r$ away from the scattering point is given by Eq. (5) in the main body of the paper, and is rewritten here for completeness [2]

$$I(r) = ev_F \left[ b^\dagger_\uparrow(-r) b^\dagger_\downarrow(-r) + b_\downarrow(-r) b_\uparrow(-r) - \mathcal{S}^*_\uparrow \mathcal{S}^*_\uparrow b^\dagger_\uparrow(r) b^\dagger_\downarrow(r) - \mathcal{S}_\downarrow \mathcal{S}_\uparrow b_\downarrow(r) b_\uparrow(r) \right]. \tag{S-1}$$

We can introduce a new set of canonical operators defined as $a_\sigma(\varepsilon) = \left[ b_\sigma(\varepsilon) + \sigma b^\dagger_{\bar\sigma}(-\varepsilon) \right]/\sqrt{2}$, in terms of which the current operator can be expressed as

$$I(r) = ev_F \Big( \sum_\sigma \left[ a^\dagger_\sigma(-r) a_\sigma(-r) - \mathrm{Re}\{\mathcal{S}_\downarrow \mathcal{S}_\uparrow\} a^\dagger_\sigma(r) a_\sigma(r) \right] + \mathrm{i}\, \mathrm{Im}\{\mathcal{S}_\downarrow \mathcal{S}_\uparrow\} \left[ a^\dagger_\uparrow(r) a^\dagger_\downarrow(r) - a_\downarrow(r) a_\uparrow(r) \right] \Big). \tag{S-2}$$

The Fermi liquid Hamiltonian (Eq. (4) in the main paper) can be expressed in this new basis. It contains elastic scattering terms (terms proportional to $\alpha_1$ and $\alpha_2$ in Eq. (S-3) below) that describe the energy dependent phase shift [3]. The inelastic effects, which arise from the quasiparticle interactions, are described by the terms proportional to $\phi_1$ and $\phi_2$. The full interacting Hamiltonian can be recast in the following form

$$\begin{aligned}
H_{\mathrm{int}} &= H_\alpha + H_\phi + \ldots \tag{S-3}\\
H_\alpha &= -\sum_\sigma \int_{\varepsilon_1,\varepsilon_2} \left[ \frac{\alpha_1}{2\pi}(\varepsilon_1+\varepsilon_2) a^\dagger_{\varepsilon_1\sigma} a_{\varepsilon_2\sigma} + \frac{\alpha_2}{4\pi}(\varepsilon_1+\varepsilon_2)^2 \left( a^\dagger_{\varepsilon_1\uparrow} a^\dagger_{\varepsilon_2\downarrow} + a_{\varepsilon_1\downarrow} a_{\varepsilon_2\uparrow} \right) \right] \\
H_\phi &= \int_{\varepsilon_1,\ldots,\varepsilon_4} \frac{\phi_1}{\pi} : a^\dagger_{\varepsilon_1\uparrow} a_{\varepsilon_2\uparrow} a^\dagger_{\varepsilon_3\downarrow} a_{\varepsilon_4\downarrow} : + \int_{\varepsilon_1,\ldots,\varepsilon_4} \frac{\phi_2}{16\pi} \Big\{ (\varepsilon_1-\varepsilon_2-\varepsilon_3+\varepsilon_4) : a^\dagger_{\varepsilon_1\uparrow} a^\dagger_{\varepsilon_2\uparrow} a^\dagger_{\varepsilon_3\downarrow} a^\dagger_{\varepsilon_4\downarrow} : \\
&\quad + (\varepsilon_1-\varepsilon_2-\varepsilon_3+\varepsilon_4) : a_{\varepsilon_1\downarrow} a_{\varepsilon_2\downarrow} a_{\varepsilon_3\uparrow} a_{\varepsilon_4\uparrow} : + 2(\varepsilon_1-\varepsilon_4) : a^\dagger_{\varepsilon_1\uparrow} a_{\varepsilon_2\uparrow} a^\dagger_{\varepsilon_3\downarrow} a^\dagger_{\varepsilon_4\uparrow} : \\
&\quad + 2(\varepsilon_4-\varepsilon_1) : a_{\varepsilon_1\uparrow} a_{\varepsilon_2\downarrow} a^\dagger_{\varepsilon_3\uparrow} a_{\varepsilon_4\uparrow} : + 2(\varepsilon_4-\varepsilon_2) : a^\dagger_{\varepsilon_1\uparrow} a^\dagger_{\varepsilon_2\downarrow} a_{\varepsilon_3\downarrow} a^\dagger_{\varepsilon_4\uparrow} : \\
&\quad + 2(\varepsilon_1-\varepsilon_3) : a_{\varepsilon_1\downarrow} a^\dagger_{\varepsilon_2\downarrow} a_{\varepsilon_3\downarrow} a_{\varepsilon_4\uparrow} : + (\varepsilon_1+\varepsilon_2-\varepsilon_3-\varepsilon_4) : a^\dagger_{\varepsilon_1\uparrow} a_{\varepsilon_2\uparrow} a_{\varepsilon_3\uparrow} a^\dagger_{\varepsilon_4\uparrow} : \\
&\quad + (-\varepsilon_1-\varepsilon_2+\varepsilon_3+\varepsilon_4) : a_{\varepsilon_1\downarrow} a^\dagger_{\varepsilon_2\uparrow} a^\dagger_{\varepsilon_3\downarrow} a_{\varepsilon_4\downarrow} : \Big\}.
\end{aligned}$$

We use perturbation theory to calculate the corrections to the average current coming from $H_{\mathrm{int}}$. They can be estimated within the Keldysh approach as

$$\langle \delta I(t) \rangle = \left\langle \mathcal{T}_\mathcal{K} I(t) e^{-\frac{i}{\hbar} \int_\mathcal{K} dt' H_{\mathrm{int}}(t')} \right\rangle, \tag{S-4}$$

where the integral is over the Keldysh contour $\mathcal{K}$. To get the corrections to the current $\sim V^3$ it is enough to expand Eq. (S-4) up to the first order in $H_{\mathrm{int}}$. The diagrams that give non-zero contributions to $\langle \delta I(t) \rangle$ are represented in Fig. S-1(a). It contains only terms coming from $\alpha_2$ and $\phi_2$, while the corrections from $\alpha_1$ and $\phi_1$ cancel out. They can be expressed in terms of the Keldysh Green's functions $i\mathcal{G}_{\sigma\sigma'}(x-x',t-t') = -\langle \mathcal{T}_\mathcal{K} a_\sigma(x,t) a^\dagger_{\sigma'}(x',t') \rangle$. In this expression, $x$ is the chiral coordinate, that can be positive or negative, while $r$ is just the separation from the junction. The non-interacting Green's functions are diagonal in spin indices and are $2 \times 2$ matrices in the Keldysh space. Their components are displayed in Fig. S-1(c). The analytical expressions in Fourier space are

$$i\mathcal{G}_{\sigma\sigma'}(\varepsilon,\omega) = i \begin{pmatrix} \mathcal{G}^t(\varepsilon,\omega) & \mathcal{G}^<(\varepsilon,\omega) \\ \mathcal{G}^>(\varepsilon,\omega) & \mathcal{G}^{\tilde t}(\varepsilon,\omega) \end{pmatrix} \delta_{\sigma\sigma'} = -\pi \begin{pmatrix} f_V^{(0)}(\varepsilon) & f_V^{(0)}(\varepsilon)+1 \\ f_V^{(0)}(\varepsilon)-1 & f_V^{(0)}(\varepsilon) \end{pmatrix} \delta(\varepsilon-\omega) \delta_{\sigma\sigma'} \tag{S-5}$$

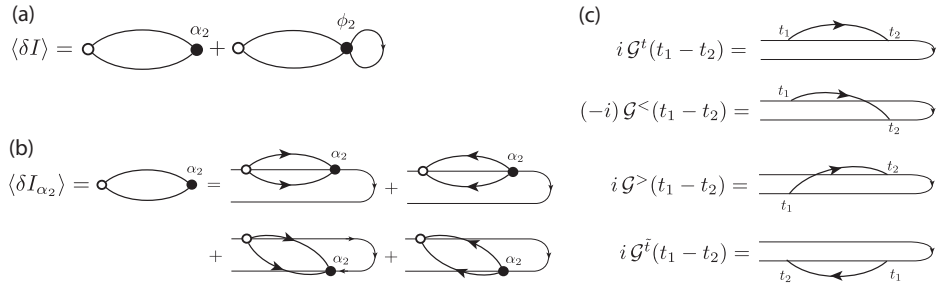

FIG. S-1: (a) Diagrams describing the corrections to the current. Only the non-vanishing contributions are displayed. The open circles represent the current vertices and the filled circles are the interacting vertices. (b) Explicit representation on the Keldysh contour for the contributions proportional to $\alpha_2$. It corresponds to the first diagram in panel (a). (c) Representation of the components of the Keldysh non-interacting Green's function. Their analytical expressions in Fourier space are given in Eq. (S-5).

with $f_V^{(0)}(\varepsilon) = 2f_V(\varepsilon) - 1$ and $f_V(\varepsilon) = f(\varepsilon - eV)$, and $f(\varepsilon)$ the regular Fermi distribution.

Inserting the current operator, Eq. (S-2), in Eq. (S-4) and using the Wick theorem, we can estimate the corrections to the average current. When computing the current average, the current vertex can be located on either branch of the Keldysh contour. Here we fix the location on the upper branch. For example, the contribution from $\alpha_2$ is detailed in Fig. S-1(b). In can be expressed as:

$$\begin{aligned}\langle \delta I_{\alpha_2}(r)\rangle &= iev_F\Big(-\frac{i}{\hbar}\Big)\Big(\frac{\alpha_2}{\pi}\Big)\mathrm{Im}\{\mathcal{S}_\uparrow \mathcal{S}_\downarrow\}\Big(\frac{1}{2\pi\hbar\, v_F}\Big)\int d\varepsilon_1 \int d\varepsilon_2 \int dt_1 \Big(\frac{\varepsilon_1-\varepsilon_2}{2}\Big)^2 \times \\ &\quad \Big\{ e^{i(k_1+k_2)x}\, i\,\mathcal{G}^t(\varepsilon_1,t_1-t)\, i\,\mathcal{G}^t(\varepsilon_2,t_1-t) - e^{-i(k_1+k_2)x}\, i\,\mathcal{G}^t(\varepsilon_1,t-t_1)\, i\,\mathcal{G}^t(\varepsilon_2,t-t_1) \\ &\quad e^{i(k_1+k_2)x}\, i\,\mathcal{G}^>(\varepsilon_1,t_1-t)\, i\,\mathcal{G}^>(\varepsilon_2,t_1-t) - e^{-i(k_1+k_2)x}\, (-i)\,\mathcal{G}^<(\varepsilon_1,t-t_1)\,(-i)\,\mathcal{G}^<(\varepsilon_2,t-t_1) \Big\} \end{aligned} \quad (S-6)$$

To evaluate Eq. (S-6), we Fourier-transform the time dependence to frequency space, and then use Eq. (S-5) to calculate it. The contributions from the time-ordered Green's functions cancel each other, and only the lesser/greater contributions remain. If we restrict ourselves to the zero temperature case, we obtain the current correction

$$\langle \delta I_{\alpha_2}(r)\rangle = V\Big(\frac{2e^2}{h}\Big) 4\alpha_2 \sin(4\delta_0)\frac{(eV)^2}{3}\,. \quad (S-7)$$

The inelastic contribution from $H_\phi$ can be computed in the same way. Similarly to what has been discussed so far, the vertex $\phi_2$ must be placed on both branches of the Keldysh contour. When explicitly represented on the Keldysh contour, the correction $\langle \delta I_{\phi_2}\rangle$ is given in terms of four diagrams. Similar to $\langle \delta I_{\alpha_2}\rangle$, the time-ordered contributions cancel, and only greater/lesser corrections remain. The evaluation of the noise proceeds along the same line. Here, we compute the noise described by the greater current correlation function: for that we have fixed the location of one current vertex on the upper branch, and the other one on the lower branch. The calculation becomes lengthly as the contributions include more diagrams, but otherwise it proceeds is a straightforward fashion.

## S-II. ALTERNATIVE APPROACH FOR THE CURRENT AND FERMI LIQUID COEFFICIENTS

The Keldysh formalism is well suited to estimate the corrections to the average current, but otherwise it is a rather heavy tool to use. In this section we present a separate route, which allows us to make quantitative estimates for the average current by including inelastic terms directly in the elastic phase shift. The starting point is the expression $\delta_\sigma(\varepsilon)$ for the energy dependent phase shift around the Fermi surface

$$\delta_\sigma(\varepsilon) = \delta + \alpha_1 \varepsilon + \alpha_2 \varepsilon^2 + \ldots. \quad (S-8)$$

At the Fermi level the phase shift is $\delta$, $\varepsilon$ is the energy measured from the Fermi level, while $\alpha_1 \sim \mathcal{O}(T_{\mathrm{FL}}^{-1})$ and $\alpha_2 \sim \mathcal{O}(T_{\mathrm{FL}}^{-2})$ are phenomenological parameters governed by the Fermi liquid energy scale $T_{\mathrm{FL}}$. In the original Kondo problem [4] $T_{\mathrm{FL}}$ is identified as the Kondo temperature $T_K$, while the phase shift at the Fermi surface is $\delta = \pi/2$. In the present situation the phase shift at the Fermi surface $\delta$ is close to $\pi/4$ and the Fermi liquid scale is $\Gamma$, i.e. the broadening parameter that describes the coupling of the dot to the normal lead. Following the strategy presented in Refs. [3, 5, 6], we can incorporate the inelastic processes described by $H_\phi$ [Eq. (S-3) in the main paper] at the Hartree level directly into the phase shift. Quite generally, these Hartree contributions are not sufficient to compute the current and we must also consider the effect of non-Hartree diagrams [5]. In the present case, there is only a single non-Hartree diagram ($\propto \phi_1^2$) at this level of perturbation theory and it vanishes. Then, $\delta_\sigma(\varepsilon)$ renormalizes accordingly

$$\delta_\sigma(\varepsilon) = \delta + \alpha_1 \varepsilon + \alpha_2 \varepsilon^2 - \sum_{\sigma' \neq \sigma} \left( \phi_1 \delta N_{0,\sigma'} + \frac{1}{2} \phi_2 \big[ \varepsilon\, \delta N_{0,\sigma'} + \delta E_{1,\sigma'} \big] \right) + \ldots, \tag{S-9}$$

where we have used the notations

$$\delta N_{0,\sigma} = \int d\varepsilon\, \delta n_\sigma(\varepsilon), \quad \delta E_{1,\sigma} = \int d\varepsilon\, \varepsilon\, \delta n_\sigma(\varepsilon), \quad \delta n_\sigma(\varepsilon) = \frac{1}{2} \left[ f_V^\sigma(\varepsilon) + f_{-V}^{-\sigma}(\varepsilon) \right] - \theta(\varepsilon_F - \varepsilon). \tag{S-10}$$

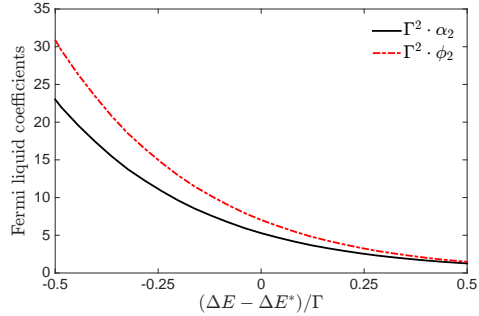

FIG. S-2: Fermi liquid coefficients $\alpha_2$ and $\phi_2$ in units of $1/\Gamma^2$ as a function of $(\Delta E - \Delta E^*)/\Gamma$.

The last identity is obtained by computing $\langle b_\sigma^\dagger(\varepsilon) b_\sigma(\varepsilon') \rangle = \frac{\delta(\varepsilon - \varepsilon')}{2} \left[ f_V^\sigma(\varepsilon) + f_{-V}^{-\sigma}(\varepsilon) \right]$. The Fermi energy $\varepsilon_F$ is arbitrary in this language - the Kondo floating argument [3, 6] is precisely based on the fact that $\delta_\sigma(\varepsilon)$ is invariant upon a shift of $\varepsilon_F$ - and it is convenient to set $\varepsilon_F = 0$ such that $\delta N_{0,\sigma}$ vanishes. In this case, $\delta E_{1,\sigma} \simeq (eV)^2/2 + (\pi T)^2/6 + \ldots$, which allows us to give the energy dependence for the phase shift of the form

$$\delta_\sigma(\varepsilon) = \delta + \alpha_1 \varepsilon + \alpha_2 \varepsilon^2 - \frac{1}{2} \phi_2 \Big[ \frac{(eV)^2}{2} + \frac{(\pi T)^2}{6} \Big] + \ldots. \tag{S-11}$$

The average current can be expressed in terms of the scattering matrix as

$$\langle I(r) \rangle = \frac{e}{h} \int d\varepsilon \left[ 1 - \mathrm{Re}\{\mathcal{S}_\downarrow(\varepsilon) \mathcal{S}_\uparrow(-\varepsilon)\} \right] \left[ f_V(\varepsilon) - f_{-V}(\varepsilon) \right], \tag{S-12}$$

Using Eq. (S-11) to construct the energy dependent scattering matrix, $\mathcal{S}_\sigma(\varepsilon) = e^{2i\delta_\sigma(\varepsilon)}$, and performing the integrals in Eq. (S-12) we recover Eq. (9) in the main paper. The noise can be computed in the same manner, *i.e.* starting with the expression (S-1) for the current operator in which the $S$ matrix includes the energy-dependent phase shift (S-11), and then performing averages over the $b$ fields in the spirit of the Landauer-Büttiker approach to the NS junction [7]. We obtain for the zero-frequency noise

$$S = \frac{4e^2}{h} \int d\varepsilon \left\{ \mathcal{T}_\varepsilon (1 - \mathcal{T}_\varepsilon) \left[ f_V(\varepsilon) - f_{-V}(\varepsilon) \right]^2 + \mathcal{T}_\varepsilon \left( f_V(\varepsilon)[1 - f_V(\varepsilon)] + f_{-V}(\varepsilon)[1 - f_{-V}(\varepsilon)] \right) \right\}, \tag{S-13}$$

written in terms of the Cooper pair energy-dependent transmission $\mathcal{T}_\varepsilon = \sin^2[\delta_\downarrow(\varepsilon) + \delta_\uparrow(\varepsilon)]$. The Keldysh perturbative result of Sec. S-I is recovered at small voltage, reproducing Eq. (10) in the main text at zero temperature.

Finally, we relate the Fermi liquid coefficients $\alpha_2$ and $\phi_2$ introduced in Eq. (S-9) to the bare parameters $\varepsilon_d$, $U$ and $\Gamma$ of the underlying single impurity Anderson model (SIAM) [8]. To conform the notations to the ones used in the main body of the paper, we relabel the single particle energy $\varepsilon_d$ with $\Delta E$. As discussed earlier, the effective Hamiltonian corresponds to the SIAM in limit when $U \to \infty$. Furthermore, the unitary limit is precisely at $\langle n_d \rangle = 1/2$, in the mixed valence regime. Similar to Ref. [6] we define $\Delta E^*$ as the single particle energy for which the occupation is $\langle n_d \rangle = 1/2$. In Fig. S-2 we plot the two dimensionless Fermi liquid parameters $\alpha_2$ and $\phi_2$. In the $U \to \infty$ limit they behave as universal functions of $(\Delta E - \Delta E^*)/\Gamma$. Using the Fermi liquid relations [6], $\alpha_2$ and $\phi_2$ can be expressed in terms of the spin and charge susceptibility derivatives [6], but the particular combination $4\alpha_2/3 - \phi_2$ that enters Eq. (8) depends only on the derivative of $\chi_c(\Delta E, \Gamma)$ with respect to $\Delta E$.

Simple formulas can in fact be written for these different quantities using the Supplemental material of Ref. [6] where the analytical expressions extracted from Bethe ansatz were simplified in the mixed valence regime. The phase shift at the Fermi energy is given by the exponentially fast converging integral

$$\delta = \frac{\pi \langle n_d \rangle}{2} = \frac{\pi}{4} - \frac{1}{2\sqrt{\pi}} \int_0^{+\infty} d\omega \, e^{-2\pi\omega} \, \mathrm{Re}\left[ i \frac{e^{-2i\pi\omega x}}{\omega} \tilde{\Gamma}\left(\frac{1}{2} + i\omega\right) \left(\frac{e}{i\omega}\right)^{i\omega} \right], \tag{S-14}$$

where we used the notation $\tilde{\Gamma}(z)$ for the gamma function and $x = (\Delta E - \Delta E^*)/\Gamma + 0.232894$. We also obtain

$$\frac{4\alpha_2}{3} - \phi_2 = -\frac{\pi \chi_c'}{3} = \frac{4\pi^{3/2}}{3\Gamma^2} \int_0^{+\infty} d\omega \, \omega \, e^{-2\pi\omega} \, \mathrm{Re}\left[ i e^{-2i\pi\omega x} \tilde{\Gamma}\left(\frac{1}{2} + i\omega\right) \left(\frac{e}{i\omega}\right)^{i\omega} \right]. \tag{S-15}$$

A similar expression exists for derivative of the spin susceptibility [6] so that $\alpha_2$ and $\phi_2$ can be evaluated independently.

## S-III. NRG CALCULATIONS AND THE HIDDEN CHARGE $U_c(1)$ SYMMETRY

In this section we provide details regarding the numerical renormalization group (NRG) [9] calculation of the matrix elements $v_\sigma^\pm$ introduced in Eq. (2) and the construction of the effective Hamiltonian (3). For that we address the problem of a superconducting (S) lead coupled to a quantum dot (QD), i.e. an S-QD setup. [13]. The underlying Hamiltonian that describes the quantum dot is the single impurity Anderson model [10],

$$H_{\mathrm{QD}} = \varepsilon_d \sum_\sigma d_\sigma^\dagger d_\sigma + U n_\uparrow n_\downarrow. \tag{S-16}$$

Here, $\varepsilon_d$ is the single particle energy and $U$ is the on-site Coulomb repulsion energy. The operator $d_\sigma^\dagger$ creates an electron with spin-$\sigma$ in the dot and $n_\sigma = d_\sigma^\dagger d_\sigma$. The BCS Hamiltonian that describes the superconducting lead can be represented by a semi-infinite Wilson chain [11],

$$H_{\mathrm{S}} = \sum_n \sum_\sigma \xi_n \Lambda^{-n/2} (f_{n\sigma}^\dagger f_{n+1\sigma} + f_{n+1\sigma}^\dagger f_{n\sigma}) + \Delta \sum_n (f_{n\uparrow}^\dagger f_{n\downarrow}^\dagger + f_{n\downarrow} f_{n\uparrow}), \tag{S-17}$$

where $f_{n\sigma}^\dagger$ is the creation operator a spin-$\sigma$ electron at site $n$ on the Wilson chain, $\xi_n$ is the nearest-neighbor hopping along the chain, and $\Lambda$ is the discretization parameter. In the second term, $\Delta$ is the superconducting gap and in the tridiagonalized Hamiltonian it is represented as an on-site pairing potential. The hybridization of the local orbital with the states in the continuum is given by

$$H_{\mathrm{tun}} = V \sum_\sigma (d_\sigma^\dagger f_{0\sigma} + f_{0\sigma}^\dagger d_\sigma). \tag{S-18}$$

The strength of the coupling $V$ between the dot and superconducting lead is assumed to be energy independent. The coupling $V$ also defines the broadening parameter $\Gamma_S = \pi \varrho V^2$, with $\varrho$ the density of states of superconductor in the normal state, which in NRG calculations is assumed to be constant, i.e. $\varrho = 1/2W$, with $2W$ the bandwidth of the conduction band. At half filling, corresponding to $\varepsilon_d = -U/2$, the total Hamiltonian describing the S-QD setup,

$$H_{\mathrm{S-QD}} = H_{\mathrm{QD}} + H_{\mathrm{S}} + H_{\mathrm{tun}} \tag{S-19}$$

is particle-hole symmetric. Apart from the spin rotation invariance, it also exhibits a hidden charge $U_c(1)$ symmetry generated by the operator

$$Q_x = \frac{1}{2}\left\{ d_\uparrow^\dagger d_\downarrow^\dagger + \sum_n (-1)^n f_{n\uparrow}^\dagger f_{n\downarrow}^\dagger + h.c. \right\}. \tag{S-20}$$

The Hamiltonian (S-19) represents the minimal model that captures the singlet-doublet transition [14] of a dot coupled to a superconducting lead. This transition can be understood as the competition between the superconducting

5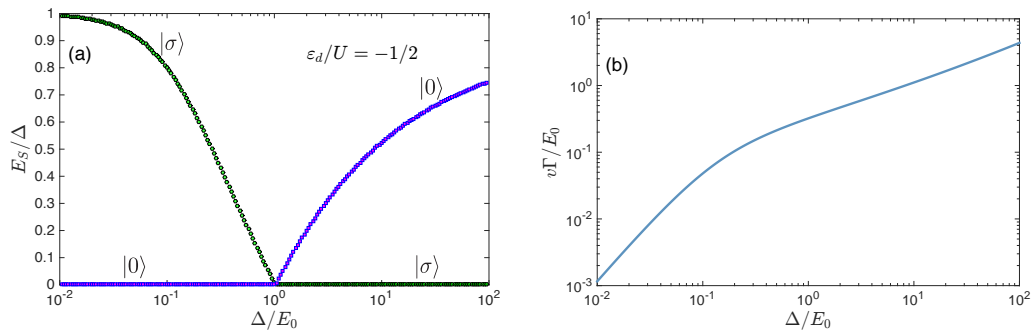

FIG. S-3: (a) The NRG results for the evolution of the Shiba states energy $E_S$ as function of $\Delta/E_0$. For $\Delta \ll E_0$, the ground state of the system is the Kondo singlet $|0\rangle$, and changes to a spin doublet $|\sigma\rangle$ for $\Delta \gg E_0$. (b) The evolution of the rescaled matrix element across the parity-changing transition has a universal behavior as a function of $\Delta/E_0$.

correlations and the Kondo screening, and takes place when $\Delta = E_0$, where $E_0$ is an energy scale of the order of the Kondo temperature $T_K$ [15]. The singlet-doublet transition is well understood if we restrict ourselves to the sub-gap states, i.e., $E < \Delta$. When $\Delta \ll E_0$, the many body ground state is a spin singlet $|0\rangle$, as the Kondo screening is strong and dominates over the superconducting correlations, while in the opposite limit, $\Delta \gg E_0$, the superconductivity wins over the Kondo correlations and the ground state becomes a doublet $|\sigma\rangle$. This is clearly visible in Fig. S-3(a) where the evolution of the sub-gap states is represented as function of $\Delta/E_0$ [12] .

The parameters $v_\sigma^\pm$ that enter Eq. (2) are the matrix elements of the dot level operators $d_\sigma^\dagger$ and $d_\sigma$ between the ground state and excited states, $v_\sigma^+ = |\langle\sigma| d_\sigma^\dagger |0\rangle|^2$ and $v_\sigma^- = |\langle\bar\sigma| d_\sigma |0\rangle|^2$. Furthermore, the electron-hole and the time-reversal symmetries guaranty that the absolute values of all the matrix elements are the same. We have explicitly checked that the matrix elements are related and satisfy the following relation: $v_\uparrow^+ = v_\downarrow^+ = v_\uparrow^- = -v_\downarrow^- = v$. The evolution of $v$ as a function of $\Delta/E_0$ is shown in Fig. S-3(b). When scaled properly it shows a universal behavior as a function of $\Delta/E_0$.

Since we are interested in the noise of the current operator across the parity changing transition, the minimal model that captures it is an effective Hamiltonian restricted to the Hilbert space of the subgap states. This is given in Eq. (3).